\documentclass[prb,aps,superscriptaddress,twocolumn,showkeys,floatfix]{revtex4-1}
\usepackage{times}
\usepackage{graphicx}

\begin{document}
\title{Order-disorder phase transition on the (100) surface of magnetite}

\author{Norman C. Bartelt}
\email{bartelt@sandia.gov}
\affiliation{Sandia National Laboratories, Livermore, CA 94550, USA}
\author{Shu Nie}
\affiliation{Sandia National Laboratories, Livermore, CA 94550, USA}
\author{Elena Starodub}
\affiliation{Sandia National Laboratories, Livermore, CA 94550, USA}
\author{Ivan Bernal}
\author{Silvia~Gallego}
\affiliation{Instituto de Ciencia de Materiales de Madrid, CSIC, Madrid 28049, Spain}
\author{Lucia~Vergara}
\affiliation{Instituto de Qu\'{\i}mica-F\'{\i}sica ``Rocasolano,'' CSIC, Madrid 28006, Spain}
\author{Kevin F. McCarty}
\affiliation{Sandia National Laboratories, Livermore, CA 94550, USA}
\author{Juan de la Figuera}
\email{juan.delafiguera@iqfr.csic.es}
\affiliation{Instituto de Qu\'{\i}mica-F\'{\i}sica ``Rocasolano,'' CSIC, Madrid 28006, Spain}

\begin{abstract}
Using low-energy electron diffraction, we show that the room-temperature $(\sqrt{2}\times\sqrt{2})R45^\circ$ reconstruction of Fe$_3$O$_4$(100) reversibly disorders at $\sim$450 $^\circ$C.  Short-range order persists above the transition, suggesting that the transition is second order and Ising-like.   We interpret the transition in terms of a model in which sub-surface Fe$^{3+}$ is replaced by Fe$^{2+}$ as the temperature is raised.   This model reproduces the structure of antiphase boundaries previously observed with STM as well as the continuous nature of the transition.  To account for the observed transition temperature, the energy cost of each charge rearrangement is 82~meV. \end{abstract}

\maketitle

Metal oxides are often useful because of their stability at high temperatures. An example is magnetite, Fe$_3$O$_4$. In catalytic applications such as the water-gas shift reaction \cite{RatnasamyCR2009} magnetite is used at temperatures between 300--500 $^\circ$C.  Furthermore, magnetite's high Curie temperature  of 580 $^\circ$C  allows spintronic applications. \cite{BibesAP2011} Since such applications frequently depend on surface properties,  a natural question arises -- how does the surface structure change with temperature?

\begin{figure}[ht]
\centerline{\includegraphics[width=0.47\textwidth]{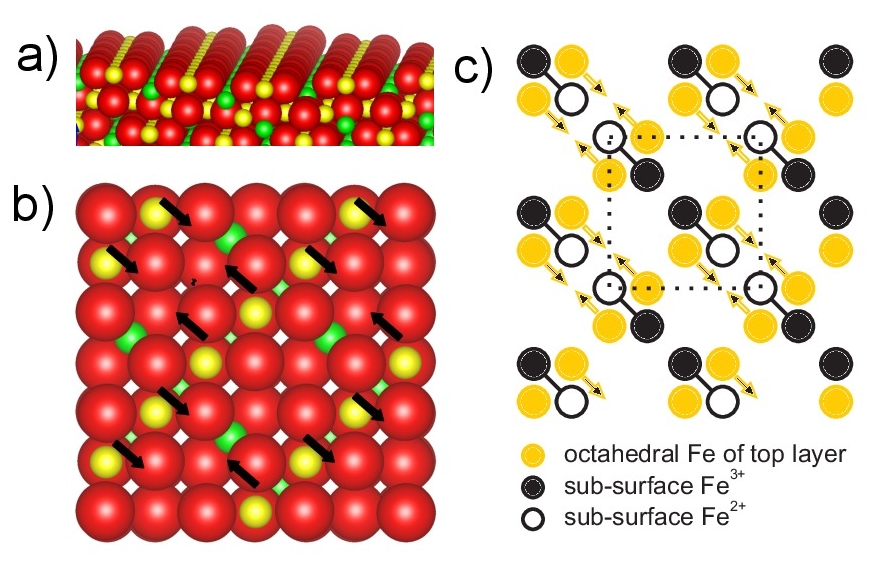}}
\caption{(color online) (a) Side-view schematic of the magnetite (100) surface terminated by octahedrally coordinated iron. The oxygen atoms, octahedral iron atoms and the tetrahedral iron atoms are colored red, yellow and green, respectively.   (b) Model of the $(\sqrt{2}\times\sqrt{2})R45^\circ$ reconstruction, where arrows indicate the displacements of the iron atoms perpendicular to their rows.  (c)  Top view of charge order in the sub-surface octahedral Fe.  The circles represent octahedral iron in the upper two layers.  The top layer iron is colored yellow, as in (b).   The black and open circles represent sub-surface Fe$^{3+}$  and Fe$^{2+}$ iron, respectively.  The $(\sqrt{2}\times\sqrt{2})R45^\circ$ cell is outlined by the dotted line. }
\label{LEEM}
\end{figure}

The room-temperature properties of magnetite's surfaces are complex. Its (100) surface has been extensively studied. \cite{DieboldSS2000,PentchevaSS2008,ChambersSS2000, BoermaJMMM2000,BoermaSS2001,WiesendangerScience1992,SchvetsPRB2002,StoltzUltra2008,ParkinsonPRB2010,ParkinsonSS2011,ShvetsPRB2006,PentchevaPRL2005,LodzianaPRL2007} Instead of the $(1 \times 1)$ bulk-like termination, it reconstructs into a structure with a larger $(\sqrt{2}\times\sqrt{2})R45^\circ$ unit cell. The atomic structure of this reconstruction has been painstakingly unraveled by density functional theory (DFT),  low-energy electron diffraction (LEED) and scanning tunneling microscopy (STM) \cite{PentchevaSS2008} -- the surface is terminated by octahedrally coordinated iron atoms arranged in rows, as shown in Fig.~\ref{LEEM}(a). The observed periodicity results from small displacements of the iron atoms perpendicular to the rows [see Fig.~\ref{LEEM}(b)].   The driving force for the reconstruction is believed to be ordering of the charge state of the iron in  octahedral sites beneath the surface.\cite{LodzianaPRL2007,GarethPRB2012} In bulk magnetite  at room temperature the average charge state of octahedral iron is +2.5$e$.  The sub-surface charge ordering involves  disproportionation of this charge into more positively and negatively charged sites.  It has been proposed \cite{LodzianaPRL2007,GarethPRB2012} that the disproportionation is greatest in the plane of octahedral iron beneath the top layer.    Fig.~\ref{LEEM}(c) sketches the proposed charge order in this subsurface layer.  (In Fig.~\ref{LEEM}(c), and in our discussion below, the two charge states in the 2nd layer are labeled by their nominal oxidation state Fe$^{3+}$ and Fe$^{2+}$, although the charges are estimated to only differ by ~0.2-0.4~{\it e} rather than {\it e}.~\cite{LodzianaPRL2007,MulakaluriPRL2009}) The top layer octahedral iron is displaced in the surface plane to decrease the distance to the nearest sub-surface Fe$^{2+}$, giving the undulating rows of surface octahedral iron observed by LEED and STM.

What might happen to such a structure when the temperature is raised?   One possibility is that the high-temperature, high-entropy surface differs significantly in stoichiometry and termination from the charge ordered $(\sqrt{2}\times\sqrt{2})R45^\circ$ phase. For example, the $(1 \times 2)$ reconstructed TiO$_2$(110) surface  has been observed to unreconstruct  at high temperature to a surface with different stoichiometry.  \cite{McCartySS2003}   In this case the transition was first order, with a discontinuous change in order and surface stoichiometry at the transition.  Another possibility is that the surface stoichiometry remains fixed and the charge order is lost.  Indeed, stoichiometric magnetite itself undergoes a first-order phase transition \footnote{The Verwey transition is first order for stoichimetric samples, while it is second order for extreme deviations of the composition (R. Arag\'{o}n, D.J. Buttrey, J.P. Shepherd, J.M. Honig, Phys. Rev. B {\bf 31}, 430 (1985)) } at $\sim$120~K, the Verwey transition,\cite{WalzJPc2002,GarciaJPc2004} which has been interpreted in terms of the disappearance of charge order.\cite{WalzJPc2002} Or the high- and low-temperature surfaces may be locally similar (with comparable charge disproportionation and iron displacements), albeit with a loss of long-range order. A continuous, second-order transition driven by configurational entropy in the charge arrangement is then a possibility.  Because the ground state shown in Fig. 1(c) is two-fold degenerate, the transition would be expected to be Ising-like.  Finally, if magnetic order were important for the surface reconstruction, one might expect that magnetite's Curie transition at 580 $^\circ$C \cite{Cornellbook} would influence the reconstruction.   

\begin{figure}[ht]
\centerline{\includegraphics[width=0.45\textwidth]{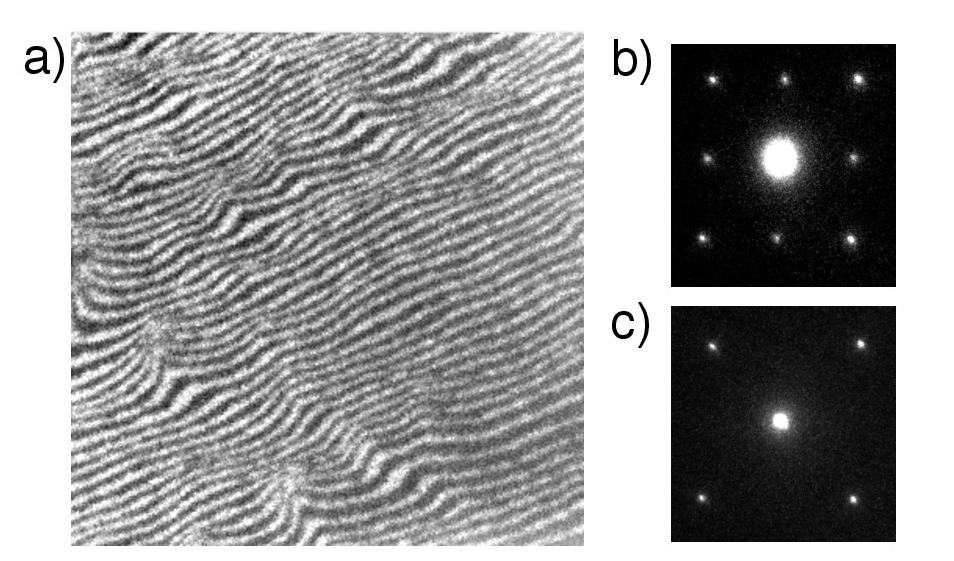}}
\caption{(a) Dark-field LEEM image acquired after several cleaning cycles. Adjacent terraces appear either black and white, a result of alternating directions of the octahedral iron rows. (b) LEED pattern of Fe$_3$O$_4$(100) at room temperature. The center spot is the (0,0) beam.   (c) LEED pattern at 488 $^\circ$C, where the reconstruction spots have disappeared. Patterns obtained with 32~eV electrons.}
\label{LEED}
\end{figure} 

Here we use LEED to study the temperature dependence of the magnetite (100) surface. We show that there is a phase transition at $\sim$450$^\circ$C. The transition is one of order/disorder, i.e., the structural motif that characterizes the surface at low temperature is still present at high temperature but is not ordered over long distances.

Two crystals of natural magnetite (100) from Mateck GmbH were examined in detail. Similar results were observed in a synthetic crystal and in a magnetite (100) film grown on SrTiO$_3$ by infrared pulsed-laser deposition.\cite{MikelASS2013} The samples were cleaned by cycles of mild sputtering (1.5 kV Ar$^+$ for 10 minutes) followed by annealing to about 600 $^\circ$C in a background of molecular oxygen at 1$\times$10$^{-6}$ Torr. Temperature was measured by a two-color infrared pyrometer. The experiments were performed in a low-energy electron microscope (LEEM\cite{altman_trends_2010}) from Elmitek GmbH, which offers several advantages over a conventional diffractometer for collecting LEED data. \cite{de_la_figuera_determiningstructure_2006,KevinSSbook} Namely, diffraction is obtained from well-defined regions that have been imaged and characterized. Also light emitted from the hot sample is isolated from the electron detector and does not degrade high-temperature measurements. 

Simply degassing the crystals in vacuum gave good LEED patterns. After cycles of sputtering and annealing no carbon or other impurities were detected on the surface by Auger electron spectroscopy. Terraces separated by steps could be observed on the surface. The dark-field LEEM image shown in Fig.~\ref{LEED}(a) was formed from a first-order diffraction beam. The Fe rows impart two-fold symmetry to a single terrace. But the row direction rotates 90$^\circ$ when crossing between adjacent terraces separated by atomic (2.1 \AA-high) steps.  So at an energy that has strong contrast between the (1,0) and (0,1) beams of a single terrace, adjacent terraces then show black and white contrast in the dark-field image. \cite{niejacs2013}   

Figure~\ref{LEED}(b) shows the LEED pattern from such a surface. The region examined, two microns in diameter, contains many atomic terraces. Thus the biaxial symmetry of an individual terrace is averaged out, giving a pattern with four-fold symmetry.  In addition to the outer $(1,0)$  spots, there are $(1/2,1/2)$ spots that correspond to a $(\sqrt{2}\times\sqrt{2})R45^\circ$ surface unit cell. The reconstruction beams become weaker upon heating the sample. At 488 $^\circ$C, the reconstruction spots are not visible [see  Fig.~\ref{LEED}(c)].  The $(\sqrt{2}\times\sqrt{2})R45^\circ$ diffraction pattern reappears upon lowering the temperature, i.e., the transition is reversible, and there is no  discernible hysteresis.   LEEM showed only a gradual change in electron reflectivity through the transition.   There was no sign of the step motion that  would signal surface mass transport due to an abrupt stoichiometry change, as was observed on TiO$_2$(110).\cite{McCartySS2003}
The transition is significantly below the Curie temperature of 580 $^\circ$C, so bulk spin order does not play a crucial role in the surface phase transition. (Spin-polarized LEEM\cite{delaFigueraUltra2013} has observed surface magnetic domains up to the bulk Curie temperature, so surface magnetic order seems to persist on the surface to the Curie point.)
\begin{figure}[h]
\centerline{\includegraphics[width=0.45\textwidth]{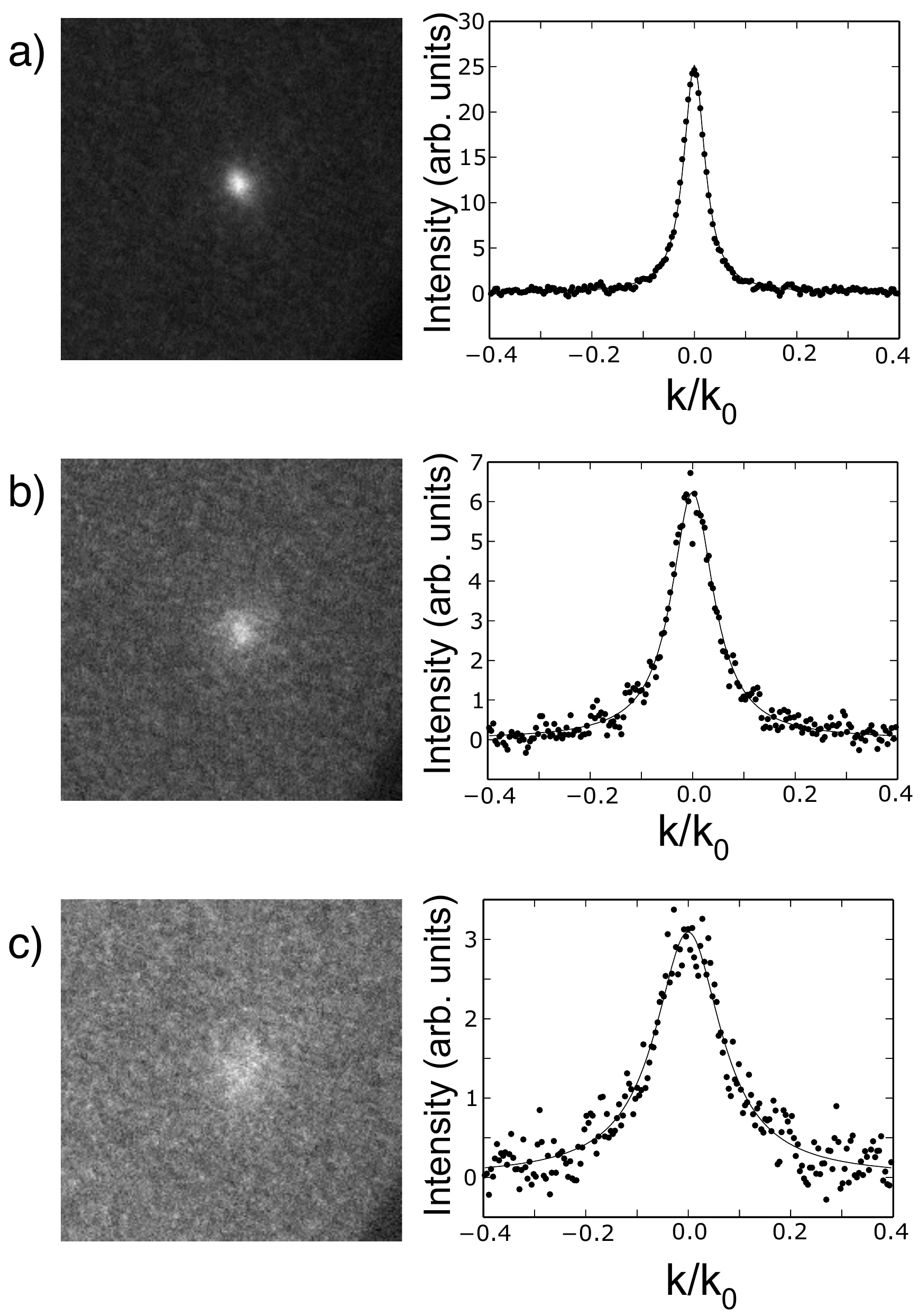}}
\caption{Profiles of the reconstruction diffracted spots at temperatures below and above the surface phase transition in 1$\times$10$^{-6}$ Torr of oxygen. (a) Image and profile of a reconstruction diffraction beam at 448 $^\circ$C.  (b) Same except for 459 $^\circ$C.  c) Same except for 465 $^\circ$C. The experimental profiles, plotted versus the in-plane diffraction vector $k$, are shown by filled circles. $k_0$ is the reciprocal lattice vector of the unreconstructed surface unit cell.  Lorentzian fits are shown by continuous lines. The size of each image is 0.8$k_0$.}
\label{profiles}
\end{figure} 

\begin{figure}[h]
\centerline{\includegraphics[width=0.6\textwidth]{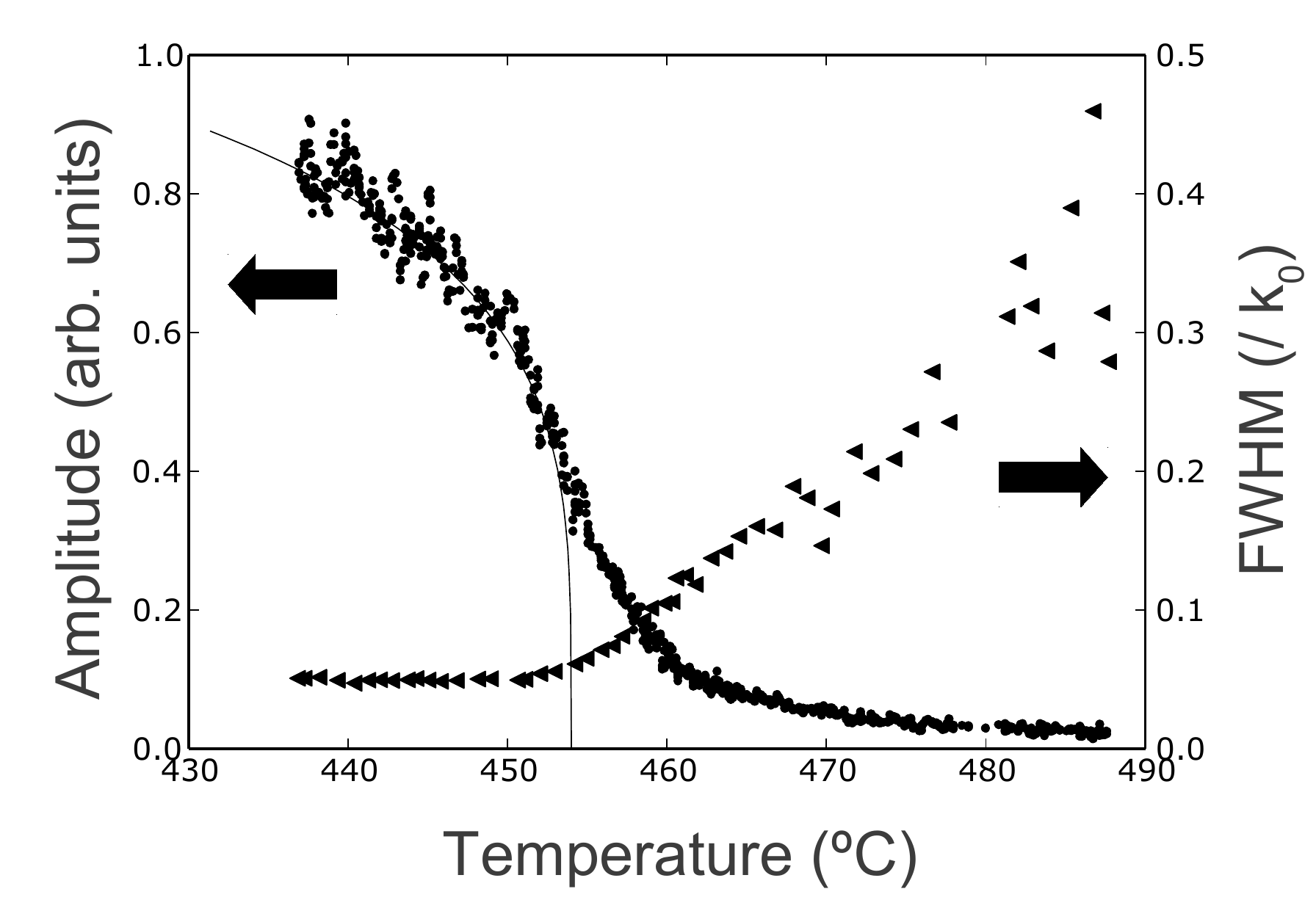}}
\caption{Amplitude (circles) and FWHM (triangles) of a $(\sqrt{2}\times\sqrt{2})R45^\circ$ reconstruction diffraction spot through the phase transition extracted from fitting to Lorentzian line shapes. The fit to the square of the order parameter of the 2D Ising model is shown as a continuous line.}
\label{fits}
\end{figure}

To gain further insight into the properties of the surface transition, we measure the profiles of the reconstruction spots through the $(\sqrt{2}\times\sqrt{2})R45^\circ \leftrightarrow (1\times1)$ transformation (see Fig.~\ref{profiles}). At all temperatures, the profiles can be reasonably fit by a Lorentzian (continuous lines in Fig.~\ref{profiles}).  The amplitude and full width at half maximum of the fitted Lorentzian is plotted in Fig.~\ref{fits}.  At low temperature the FWHM is limited by instrumental resolution.  This is consistent with long-range $(\sqrt{2}\times\sqrt{2})R45^\circ$ order.  Above $\sim$450 $^\circ$C, the width of the reconstruction beam increases. This large increase, to 10 times the instrumental resolution, is characteristic of  scattering above the critical temperature $T_c$ of a second-order phase transition when long-range order is replaced by short-range order with a finite correlation length.  (In the limit of perfect instrumental resolution, the FWHM is inversely proportional to the correlation length.) Since the ground state of the reconstruction is two-fold degenerate, we expect its second-order transition to be reasonably described by a two-dimensional  Ising model. \cite{RoelofsHSS1996}  Beneath $T_c$  the temperature dependence of the spot intensity due to long-range order is proportional to the square of the order parameter of the transition. The continuous line in Fig.~\ref{fits} fits this diffraction data to the square of the Ising order parameter $M$:
\begin{equation}
M = \left( 1- \sinh (\ln((1+\sqrt{2})T/T_c)^{-4}\right)^\frac{1}{8},
\end{equation}
where $T_c$ is the transition temperature.  The agreement between the measured diffraction intensity and the 2D Ising model is  good. We estimate $T_c$ to be 454~$^\circ$C.  Furthermore, the approximately linear increase in the FWHM (i.e., the inverse correlation length) is also consistent with the Ising model. \cite{RoelofsHSS1996} Thus, the 2D Ising model provides a reasonable description of the temperature dependence of the $(\sqrt{2}\times\sqrt{2})R45^\circ \leftrightarrow (1\times1)$ transition. \footnote{Lack of data very near to $T_c$ precludes making useful estimates of critical exponents.}      We did not observe time dependent fluctuations in the LEEM image intensity which can be caused by critical fluctuations  near $T_c$ (Ref. \onlinecite{trompprl1996}], presumably because they were much faster than our image acquisition time of 0.03 s.

The transition temperature was the same in vacuum or 1$\times$10$^{-6}$ Torr of oxygen. While we cannot exclude a dependence of the transition temperature on the crystal's bulk stoichiometry, we note that our natural crystal and a synthetic crystal had the same transition temperature within a few tens of degrees. 

\begin{figure}
\centerline{\includegraphics[width=0.40\textwidth]{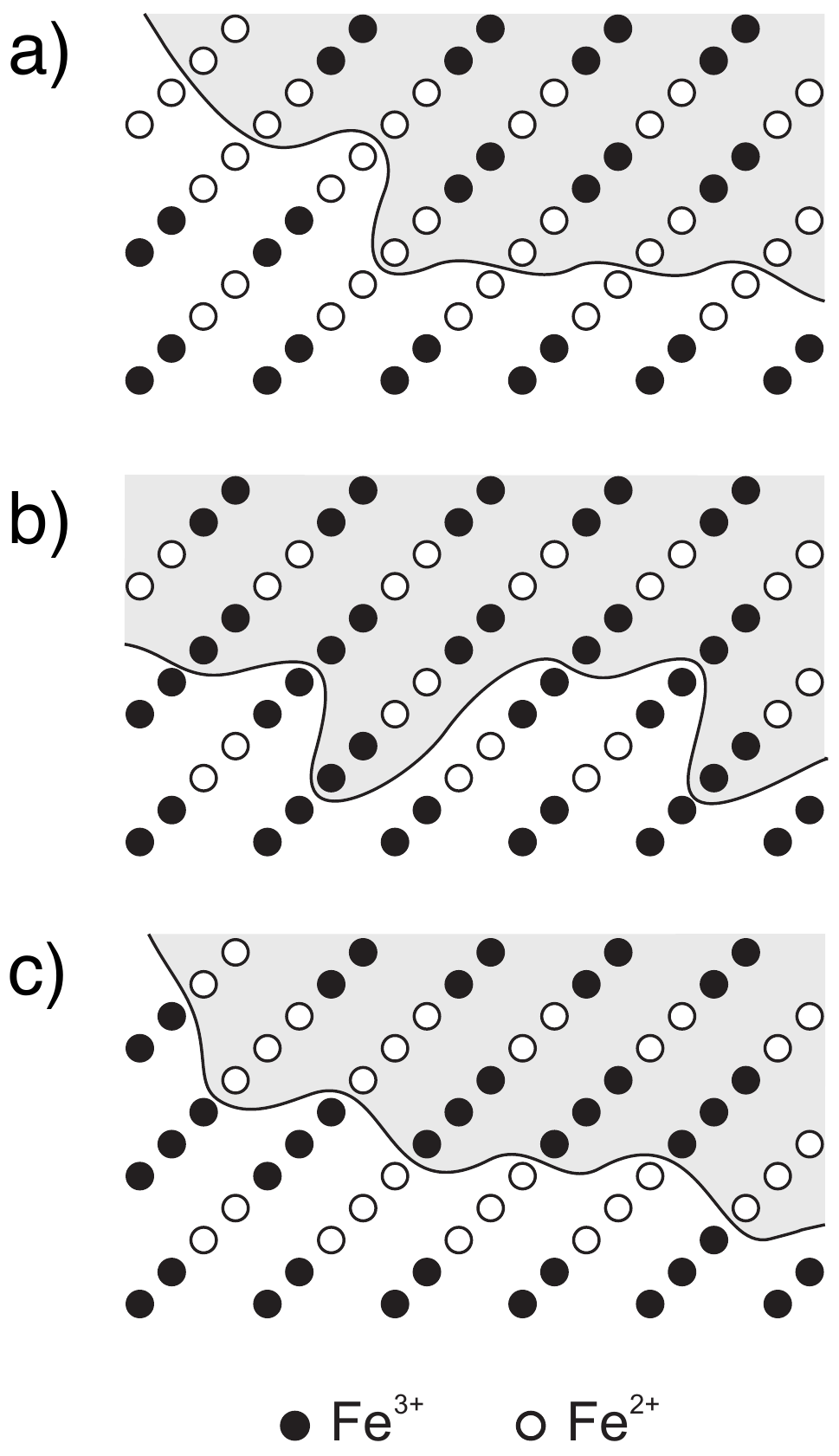}}
\caption{(a) The structure of  anti-phase boundaries in the $(\sqrt{2}\times\sqrt{2})R45^\circ$ reconstruction reported in Ref. [\onlinecite{GarethPRB2012}].  The configurations drawn in (b) and (c) were not observed.}
\label{antiphase}
\end{figure}

\begin{figure}
\centerline{\includegraphics[width=0.40\textwidth]{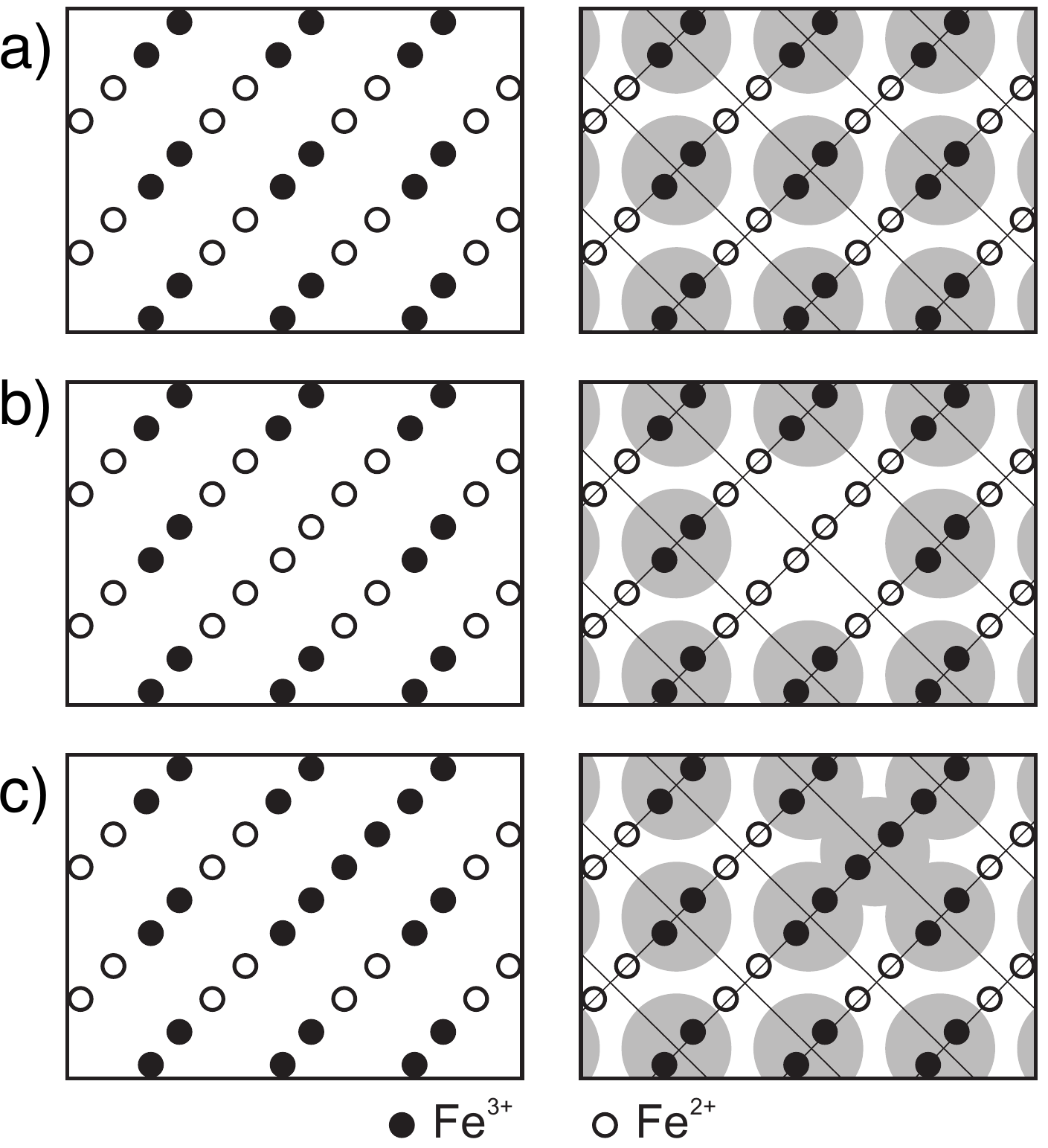}}
\caption{Mapping configurations of charges (left hand panels) onto configurations of occupied sites  of the hard square model (large circles in the right hand panels).  (a) Ground state.  (b) The lowest-energy allowed excitation of the ground state that replaces a pair of Fe$^{3+}$ with Fe$^{2+}$ and corresponds to creating a  vacancy in the ordered state of the  hard square model.  (c) An forbidden configuration in which Fe$^{3+}$ replaces Fe$^{2+}$.  In the hard square model this is not allowed because nearest neighbor sites in the square lattice are simultaneously occupied.}
\label{excitation}
\end{figure}

Given the continuous nature of the transition and the absence of any direct evidence for a stoichiometry change,  it is likely that the transition is driven by configurational entropy  in the charge order.   To interpret the observed $T_c$  then requires understanding the low-energy excitations of the charge-order pattern.  The  structure of anti-phase domain boundaries deduced from STM images by Parkinson {\it et al.} \cite{GarethPRB2012} provide significant information about  these excitations.   They find that the boundary between the two degenerate ground states always consists of four adjacent  Fe$^{2+}$ in a row, as shown schematically in Fig. \ref{antiphase}(a).   Other possible types of antiphase domains with four  Fe$^{3+}$ in a row or three Fe$^{2+}$ and Fe$^{3+}$ in a row (Fig. \ref{antiphase}(b) and (c), respectively) were not observed.  Thus the phase boundaries appear to be relatively rich in sub-surface Fe$^{2+}$.
By comparing allowed and disallowed configurations, we can hypothesize two microscopic rules governing the observed antiphase boundaries: 1) Fe$^{2+}$ and Fe$^{3+}$ are always arranged in pairs because anti-phase boundaries such as Fig. \ref{antiphase}(c) were not observed,  2) The four nearest neighbor pairs of an Fe$^{3+}$ pair are always Fe$^{2+}$ pairs because phase boundaries such as   Fig. \ref{antiphase}(b) were not observed.
These rules would imply that the lowest energy excitation of the $(\sqrt{2}\times\sqrt{2})R45^\circ$ ground state is a replacement of an Fe$^{3+}$ pair by an Fe$^{2+}$ pair, as shown in the left hand panels of Figs. \ref{excitation}(a) and (b).
Thus as the temperature is raised, sub-surface Fe$^{2+}$ defects will be gradually created.   When the density of these defects becomes  large, anti phase boundaries can form.  Eventually, long-range $(\sqrt{2}\times\sqrt{2})R45^\circ$ order will be lost as the anti-phase boundaries  proliferate.   To estimate this  critical defect density, and to give $T_c$ in terms of the energy $E$ of the lowest  energy excitation we can map our problem onto the hard square lattice gas model of statistical mechanics as shown in  the right hand panels of Fig. \ref{excitation}.  The hard square model assumes that  sites of a square lattice can be occupied only if nearest neighbor sites are unoccupied.   If one identifies Fe$^{3+}$ pairs with occupied lattice sites, then the hard square lattice gas rule upholds the rule 2,  the maximum coverage of occupied sites is 1/2 (in the $(\sqrt{2}\times\sqrt{2})R45^\circ$ state), as in Fig. \ref{excitation}(a). Occupation of nearest neighbor sites, which correspond to adding Fe$^{3+}$ charges to the ordered state as in Fig. \ref{excitation}(c), is not allowed.  As shown in Fig. \ref{excitation}(b), the lowest energy excitation  is equivalent to forming a vacancy in the hard square model.   

The$(\sqrt{2}\times\sqrt{2})R45^\circ$ ordered state of the hard square model  is known to disorder in an Ising-like second-order transition when the coverage of occupied sites drops from 0.5 to $\sim0.368$.  (See, e.g., Ref. [\onlinecite{kinzel1981}].) To relate this transition to the observed disordering of the magnetite's reconstruction and the energy $E$ to create a vacancy, we note that the partition function $Z$ of our problem, taking the zero in energy as the $(\sqrt{2}\times\sqrt{2})R45^\circ$ ground state, is 
\begin{equation}
Z= \sum e^{\left( {N \over 2}-n\right) E /kT} = e^{-NE/2kT}\sum z^n,
\end{equation}                      
where the sum is over all allowed configuration, $N$ is the total number of sites, $n$ is the number of occupied sites (i.e. Fe$^{3+}$ pairs)  and  $z=e^{E/kT}$.    The order-disorder transition occurs when $\ln(z)=E/kT_c \approx 1.33$.\cite{kinzel1981}   From the measured $T_c =454$~$^\circ$C the energy of the rearrangement (changing two  Fe$^{3+}$ to two Fe$^{2+}$) is  thus predicted to be 82 meV.   This number could provide a check to density functional theory models of the stability of the reconstruction.\footnote{The complications of enforcing charge states in this system make theoretical estimates difficult.    Calculations in Ref. [\onlinecite{GarethPRB2012}]  suggest that replacing one Fe$^{3+}$ with Fe$^{2+}$ costs 55 meV, which would be of the same order as we predict, but since the details of the configuration studied is not given, it is difficult to directly compare with our model.}  Preliminary DFT calculations \footnote{I. Bernal and S. Gallego, to be published.} suggest that completely removing the charge order only changes the energy per unit cell by a few $kT$ per unit cell, lending support to the plausibility that the transition is driven by configurational entropy of the charge order and not changes in stoichiometry. 

Another prediction of this model is that the second layer octahedral iron would gradually become more reduced as $T$ is raised.   Perhaps this change would be compensated by a more oxidized top layer (which is already known to be Fe$^{3+}$ rich at room temperature \cite{ChambersSS2000,ParkinsonPRB2013}).

As mentioned, the charge-ordered $(\sqrt{2}\times\sqrt{2})R45^\circ$  reconstruction has been proposed as the surface equivalent of the low-temperature monoclinic phase of bulk magnetite. \cite{LodzianaPRL2007}   The charge order in the monoclinic phase is complex, \cite{SennNature2012}  with many inequivalent states of the octahedral iron.  Therefore, it s likely that models such as in Fig. 1(c), where disproportionation is restricted to two charge states in a single octahedral plane, are a considerable oversimplification.   Thus, there could well be other degrees of freedom and sources of entropy that would influence the transition temperature.  More detailed calculations of the nature of the surface charge order are clearly needed to address this question.    The analogy with the monoclinic phase also
 begs the question of whether the transition we observe at high temperature, $(\sqrt{2}\times\sqrt{2})R45^\circ \leftrightarrow (1\times1)$, can be interpreted as a surface analog of the Verwey transition.\cite{LodzianaPRL2007} Strictly, the bulk Verwey transition is a first-order transition that involves a substantial structural change, a small change in the charge order and limited changes in the spin order. So from this point of view, our transition is not simply a `surface' Verwey transition. Nevertheless, our model  suggests that an interplay between charge order and atomic structure is responsible for the surface phase transition.

In summary, we have experimentally observed that the magnetite (100) surface undergoes a $(\sqrt{2}\times\sqrt{2})R45^\circ \leftrightarrow (1\times1)$ phase transition.  Motivated by the structure of antiphase boundaries observed in Ref. [\onlinecite{GarethPRB2012}] we speculate that the transition is caused by a gradual reduction of sub-surface octahedral iron as the temperature is raised.  

\section*{Acknowledgments}

We thank Prof. Gareth Parkinson for lending us a synthetic magnetite crystal and Dr. Mikel Sanz for providing the magnetite thin film. This research was supported by the Office of Basic Energy Sciences, Division of Materials and Engineering Sciences, U.~S. Department of Energy under Contract No. DE-AC04-94AL85000 and by the goverment of Spain through Project No.~MAT2009-14578-C03-01/03 and MAT2012-38045-C04-01/04. LV and IB thanks the Spanish CSIC for support through a JAE-Doc contract.

\bibliography{magnetite}

\begin{thebibliography}{36}%
\makeatletter
\providecommand \@ifxundefined [1]{%
 \@ifx{#1\undefined}
}%
\providecommand \@ifnum [1]{%
 \ifnum #1\expandafter \@firstoftwo
 \else \expandafter \@secondoftwo
 \fi
}%
\providecommand \@ifx [1]{%
 \ifx #1\expandafter \@firstoftwo
 \else \expandafter \@secondoftwo
 \fi
}%
\providecommand \natexlab [1]{#1}%
\providecommand \enquote  [1]{``#1''}%
\providecommand \bibnamefont  [1]{#1}%
\providecommand \bibfnamefont [1]{#1}%
\providecommand \citenamefont [1]{#1}%
\providecommand \href@noop [0]{\@secondoftwo}%
\providecommand \href [0]{\begingroup \@sanitize@url \@href}%
\providecommand \@href[1]{\@@startlink{#1}\@@href}%
\providecommand \@@href[1]{\endgroup#1\@@endlink}%
\providecommand \@sanitize@url [0]{\catcode `\\12\catcode `\$12\catcode
  `\&12\catcode `\#12\catcode `\^12\catcode `\_12\catcode `\%12\relax}%
\providecommand \@@startlink[1]{}%
\providecommand \@@endlink[0]{}%
\providecommand \url  [0]{\begingroup\@sanitize@url \@url }%
\providecommand \@url [1]{\endgroup\@href {#1}{\urlprefix }}%
\providecommand \urlprefix  [0]{URL }%
\providecommand \Eprint [0]{\href }%
\providecommand \doibase [0]{http://dx.doi.org/}%
\providecommand \selectlanguage [0]{\@gobble}%
\providecommand \bibinfo  [0]{\@secondoftwo}%
\providecommand \bibfield  [0]{\@secondoftwo}%
\providecommand \translation [1]{[#1]}%
\providecommand \BibitemOpen [0]{}%
\providecommand \bibitemStop [0]{}%
\providecommand \bibitemNoStop [0]{.\EOS\space}%
\providecommand \EOS [0]{\spacefactor3000\relax}%
\providecommand \BibitemShut  [1]{\csname bibitem#1\endcsname}%
\let\auto@bib@innerbib\@empty
\bibitem [{\citenamefont {Ratnasamy}\ and\ \citenamefont
  {Wagner}(2009)}]{RatnasamyCR2009}%
  \BibitemOpen
  \bibfield  {author} {\bibinfo {author} {\bibfnamefont {C.}~\bibnamefont
  {Ratnasamy}}\ and\ \bibinfo {author} {\bibfnamefont {J.~P.}\ \bibnamefont
  {Wagner}},\ }\href {\doibase 10.1080/01614940903048661} {\bibfield  {journal}
  {\bibinfo  {journal} {Catal. Rev.}\ }\textbf {\bibinfo {volume} {51}},\
  \bibinfo {pages} {325} (\bibinfo {year} {2009})}\BibitemShut {NoStop}%
\bibitem [{\citenamefont {Bibes}\ \emph {et~al.}(2011)\citenamefont {Bibes},
  \citenamefont {Villegas},\ and\ \citenamefont
  {Barth{\'e}l{\'e}my}}]{BibesAP2011}%
  \BibitemOpen
  \bibfield  {author} {\bibinfo {author} {\bibfnamefont {M.}~\bibnamefont
  {Bibes}}, \bibinfo {author} {\bibfnamefont {J.~E.}\ \bibnamefont {Villegas}},
  \ and\ \bibinfo {author} {\bibfnamefont {A.}~\bibnamefont
  {Barth{\'e}l{\'e}my}},\ }\href {\doibase 10.1080/00018732.2010.534865}
  {\bibfield  {journal} {\bibinfo  {journal} {Adv. Phys.}\ }\textbf {\bibinfo
  {volume} {60}},\ \bibinfo {pages} {5} (\bibinfo {year} {2011})}\BibitemShut
  {NoStop}%
\bibitem [{\citenamefont {Stanka}\ \emph {et~al.}(2000)\citenamefont {Stanka},
  \citenamefont {Hebenstreit}, \citenamefont {Diebold},\ and\ \citenamefont
  {Chambers}}]{DieboldSS2000}%
  \BibitemOpen
  \bibfield  {author} {\bibinfo {author} {\bibfnamefont {B.}~\bibnamefont
  {Stanka}}, \bibinfo {author} {\bibfnamefont {W.}~\bibnamefont {Hebenstreit}},
  \bibinfo {author} {\bibfnamefont {U.}~\bibnamefont {Diebold}}, \ and\
  \bibinfo {author} {\bibfnamefont {S.}~\bibnamefont {Chambers}},\ }\href
  {\doibase 10.1016/S0039-6028(99)01182-6} {\bibfield  {journal} {\bibinfo
  {journal} {Surf. Sci.}\ }\textbf {\bibinfo {volume} {448}},\ \bibinfo {pages}
  {49} (\bibinfo {year} {2000})}\BibitemShut {NoStop}%
\bibitem [{\citenamefont {Pentcheva}\ \emph {et~al.}(2008)\citenamefont
  {Pentcheva}, \citenamefont {Moritz}, \citenamefont {Rundgren}, \citenamefont
  {Frank}, \citenamefont {Schrupp},\ and\ \citenamefont
  {Scheffler}}]{PentchevaSS2008}%
  \BibitemOpen
  \bibfield  {author} {\bibinfo {author} {\bibfnamefont {R.}~\bibnamefont
  {Pentcheva}}, \bibinfo {author} {\bibfnamefont {W.}~\bibnamefont {Moritz}},
  \bibinfo {author} {\bibfnamefont {J.}~\bibnamefont {Rundgren}}, \bibinfo
  {author} {\bibfnamefont {S.}~\bibnamefont {Frank}}, \bibinfo {author}
  {\bibfnamefont {D.}~\bibnamefont {Schrupp}}, \ and\ \bibinfo {author}
  {\bibfnamefont {M.}~\bibnamefont {Scheffler}},\ }\href {\doibase
  16/j.susc.2008.01.006} {\bibfield  {journal} {\bibinfo  {journal} {Surf.
  Sci.}\ }\textbf {\bibinfo {volume} {602}},\ \bibinfo {pages} {1299} (\bibinfo
  {year} {2008})}\BibitemShut {NoStop}%
\bibitem [{\citenamefont {Chambers}\ \emph {et~al.}(2000)\citenamefont
  {Chambers}, \citenamefont {Thevuthasan},\ and\ \citenamefont
  {Joyce}}]{ChambersSS2000}%
  \BibitemOpen
  \bibfield  {author} {\bibinfo {author} {\bibfnamefont {S.}~\bibnamefont
  {Chambers}}, \bibinfo {author} {\bibfnamefont {S.}~\bibnamefont
  {Thevuthasan}}, \ and\ \bibinfo {author} {\bibfnamefont {S.}~\bibnamefont
  {Joyce}},\ }\href {\doibase 10.1016/S0039-6028(00)00230-2} {\bibfield
  {journal} {\bibinfo  {journal} {Surf. Sci.}\ }\textbf {\bibinfo {volume}
  {450}},\ \bibinfo {pages} {L273} (\bibinfo {year} {2000})}\BibitemShut
  {NoStop}%
\bibitem [{\citenamefont {Mijiritskii}\ \emph {et~al.}(2000)\citenamefont
  {Mijiritskii}, \citenamefont {Langelaar},\ and\ \citenamefont
  {Boerma}}]{BoermaJMMM2000}%
  \BibitemOpen
  \bibfield  {author} {\bibinfo {author} {\bibfnamefont {A.}~\bibnamefont
  {Mijiritskii}}, \bibinfo {author} {\bibfnamefont {M.}~\bibnamefont
  {Langelaar}}, \ and\ \bibinfo {author} {\bibfnamefont {D.}~\bibnamefont
  {Boerma}},\ }\href {\doibase 10.1016/S0304-8853(99)00747-7} {\bibfield
  {journal} {\bibinfo  {journal} {J. Mag. Mag. Mat.}\ }\textbf {\bibinfo
  {volume} {211}},\ \bibinfo {pages} {278} (\bibinfo {year}
  {2000})}\BibitemShut {NoStop}%
\bibitem [{\citenamefont {Mijiritskii}\ and\ \citenamefont
  {Boerma}(2001)}]{BoermaSS2001}%
  \BibitemOpen
  \bibfield  {author} {\bibinfo {author} {\bibfnamefont {A.}~\bibnamefont
  {Mijiritskii}}\ and\ \bibinfo {author} {\bibfnamefont {D.}~\bibnamefont
  {Boerma}},\ }\href {\doibase 10.1016/S0039-6028(01)01064-0} {\bibfield
  {journal} {\bibinfo  {journal} {Surf. Sci.}\ }\textbf {\bibinfo {volume}
  {486}},\ \bibinfo {pages} {73} (\bibinfo {year} {2001})}\BibitemShut
  {NoStop}%
\bibitem [{\citenamefont {Wiesendanger}\ \emph {et~al.}(1992)\citenamefont
  {Wiesendanger}, \citenamefont {Shvets}, \citenamefont {B{\"u}rgler},
  \citenamefont {Tarrach}, \citenamefont {G\"{u}ntherodt}, \citenamefont
  {Coey},\ and\ \citenamefont {Gr\"{a}ser}}]{WiesendangerScience1992}%
  \BibitemOpen
  \bibfield  {author} {\bibinfo {author} {\bibfnamefont {R.}~\bibnamefont
  {Wiesendanger}}, \bibinfo {author} {\bibfnamefont {I.~V.}\ \bibnamefont
  {Shvets}}, \bibinfo {author} {\bibfnamefont {D.}~\bibnamefont {B{\"u}rgler}},
  \bibinfo {author} {\bibfnamefont {G.}~\bibnamefont {Tarrach}}, \bibinfo
  {author} {\bibfnamefont {H.~J.}\ \bibnamefont {G\"{u}ntherodt}}, \bibinfo
  {author} {\bibfnamefont {J.~M.~D.}\ \bibnamefont {Coey}}, \ and\ \bibinfo
  {author} {\bibfnamefont {S.}~\bibnamefont {Gr\"{a}ser}},\ }\href {\doibase
  10.1126/science.255.5044.583} {\bibfield  {journal} {\bibinfo  {journal}
  {Science}\ }\textbf {\bibinfo {volume} {255}},\ \bibinfo {pages} {583}
  (\bibinfo {year} {1992})}\BibitemShut {NoStop}%
\bibitem [{\citenamefont {Mariotto}\ \emph {et~al.}(2002)\citenamefont
  {Mariotto}, \citenamefont {Murphy},\ and\ \citenamefont
  {Shvets}}]{SchvetsPRB2002}%
  \BibitemOpen
  \bibfield  {author} {\bibinfo {author} {\bibfnamefont {G.}~\bibnamefont
  {Mariotto}}, \bibinfo {author} {\bibfnamefont {S.}~\bibnamefont {Murphy}}, \
  and\ \bibinfo {author} {\bibfnamefont {I.~V.}\ \bibnamefont {Shvets}},\
  }\href {\doibase 10.1103/PhysRevB.66.245426} {\bibfield  {journal} {\bibinfo
  {journal} {Phys. Rev. B}\ }\textbf {\bibinfo {volume} {66}},\ \bibinfo
  {pages} {245426} (\bibinfo {year} {2002})}\BibitemShut {NoStop}%
\bibitem [{\citenamefont {Stoltz}\ \emph {et~al.}(2008)\citenamefont {Stoltz},
  \citenamefont {{\"O}nsten}, \citenamefont {Karlsson},\ and\ \citenamefont
  {G{\"o}thelid}}]{StoltzUltra2008}%
  \BibitemOpen
  \bibfield  {author} {\bibinfo {author} {\bibfnamefont {D.}~\bibnamefont
  {Stoltz}}, \bibinfo {author} {\bibfnamefont {A.}~\bibnamefont {{\"O}nsten}},
  \bibinfo {author} {\bibfnamefont {U.}~\bibnamefont {Karlsson}}, \ and\
  \bibinfo {author} {\bibfnamefont {M.}~\bibnamefont {G{\"o}thelid}},\ }\href
  {\doibase 10.1016/j.ultramic.2007.08.010} {\bibfield  {journal} {\bibinfo
  {journal} {Ultramicroscopy}\ }\textbf {\bibinfo {volume} {108}},\ \bibinfo
  {pages} {540} (\bibinfo {year} {2008})}\BibitemShut {NoStop}%
\bibitem [{\citenamefont {Parkinson}\ \emph {et~al.}(2010)\citenamefont
  {Parkinson}, \citenamefont {Mulakaluri}, \citenamefont {Losovyj},
  \citenamefont {Jacobson}, \citenamefont {Pentcheva},\ and\ \citenamefont
  {Diebold}}]{ParkinsonPRB2010}%
  \BibitemOpen
  \bibfield  {author} {\bibinfo {author} {\bibfnamefont {G.~S.}\ \bibnamefont
  {Parkinson}}, \bibinfo {author} {\bibfnamefont {N.}~\bibnamefont
  {Mulakaluri}}, \bibinfo {author} {\bibfnamefont {Y.}~\bibnamefont {Losovyj}},
  \bibinfo {author} {\bibfnamefont {P.}~\bibnamefont {Jacobson}}, \bibinfo
  {author} {\bibfnamefont {R.}~\bibnamefont {Pentcheva}}, \ and\ \bibinfo
  {author} {\bibfnamefont {U.}~\bibnamefont {Diebold}},\ }\href {\doibase
  10.1103/PhysRevB.82.125413} {\bibfield  {journal} {\bibinfo  {journal} {Phys.
  Rev. B}\ }\textbf {\bibinfo {volume} {82}},\ \bibinfo {pages} {125413}
  (\bibinfo {year} {2010})}\BibitemShut {NoStop}%
\bibitem [{\citenamefont {Parkinson}\ \emph {et~al.}(2011)\citenamefont
  {Parkinson}, \citenamefont {Novotn\'{y}}, \citenamefont {Jacobson},
  \citenamefont {Schmid},\ and\ \citenamefont {Diebold}}]{ParkinsonSS2011}%
  \BibitemOpen
  \bibfield  {author} {\bibinfo {author} {\bibfnamefont {G.~S.}\ \bibnamefont
  {Parkinson}}, \bibinfo {author} {\bibfnamefont {Z.}~\bibnamefont
  {Novotn\'{y}}}, \bibinfo {author} {\bibfnamefont {P.}~\bibnamefont
  {Jacobson}}, \bibinfo {author} {\bibfnamefont {M.}~\bibnamefont {Schmid}}, \
  and\ \bibinfo {author} {\bibfnamefont {U.}~\bibnamefont {Diebold}},\ }\href
  {\doibase 10.1016/j.susc.2011.05.018} {\bibfield  {journal} {\bibinfo
  {journal} {Surf. Sci.}\ }\textbf {\bibinfo {volume} {605}},\ \bibinfo {pages}
  {L42} (\bibinfo {year} {2011})}\BibitemShut {NoStop}%
\bibitem [{\citenamefont {Jordan}\ \emph {et~al.}(2006)\citenamefont {Jordan},
  \citenamefont {Cazacu}, \citenamefont {Manai}, \citenamefont {Ceballos},
  \citenamefont {Murphy},\ and\ \citenamefont {Shvets}}]{ShvetsPRB2006}%
  \BibitemOpen
  \bibfield  {author} {\bibinfo {author} {\bibfnamefont {K.}~\bibnamefont
  {Jordan}}, \bibinfo {author} {\bibfnamefont {A.}~\bibnamefont {Cazacu}},
  \bibinfo {author} {\bibfnamefont {G.}~\bibnamefont {Manai}}, \bibinfo
  {author} {\bibfnamefont {S.~F.}\ \bibnamefont {Ceballos}}, \bibinfo {author}
  {\bibfnamefont {S.}~\bibnamefont {Murphy}}, \ and\ \bibinfo {author}
  {\bibfnamefont {I.~V.}\ \bibnamefont {Shvets}},\ }\href {\doibase
  10.1103/PhysRevB.74.085416} {\bibfield  {journal} {\bibinfo  {journal} {Phys.
  Rev. B}\ }\textbf {\bibinfo {volume} {74}},\ \bibinfo {pages} {085416}
  (\bibinfo {year} {2006})}\BibitemShut {NoStop}%
\bibitem [{\citenamefont {Pentcheva}\ \emph {et~al.}(2005)\citenamefont
  {Pentcheva}, \citenamefont {Wendler}, \citenamefont {Meyerheim},
  \citenamefont {Moritz}, \citenamefont {Jedrecy},\ and\ \citenamefont
  {Scheffler}}]{PentchevaPRL2005}%
  \BibitemOpen
  \bibfield  {author} {\bibinfo {author} {\bibfnamefont {R.}~\bibnamefont
  {Pentcheva}}, \bibinfo {author} {\bibfnamefont {F.}~\bibnamefont {Wendler}},
  \bibinfo {author} {\bibfnamefont {H.~L.}\ \bibnamefont {Meyerheim}}, \bibinfo
  {author} {\bibfnamefont {W.}~\bibnamefont {Moritz}}, \bibinfo {author}
  {\bibfnamefont {N.}~\bibnamefont {Jedrecy}}, \ and\ \bibinfo {author}
  {\bibfnamefont {M.}~\bibnamefont {Scheffler}},\ }\href {\doibase
  10.1103/PhysRevLett.94.126101} {\bibfield  {journal} {\bibinfo  {journal}
  {Phys. Rev. Lett.}\ }\textbf {\bibinfo {volume} {94}},\ \bibinfo {pages}
  {126101} (\bibinfo {year} {2005})}\BibitemShut {NoStop}%
\bibitem [{\citenamefont {Lodziana}(2007)}]{LodzianaPRL2007}%
  \BibitemOpen
  \bibfield  {author} {\bibinfo {author} {\bibfnamefont {Z.}~\bibnamefont
  {Lodziana}},\ }\href {\doibase 10.1103/PhysRevLett.99.206402} {\bibfield
  {journal} {\bibinfo  {journal} {Phys. Rev. Lett.}\ }\textbf {\bibinfo
  {volume} {99}},\ \bibinfo {pages} {206402} (\bibinfo {year}
  {2007})}\BibitemShut {NoStop}%
\bibitem [{\citenamefont {Parkinson}\ \emph {et~al.}(2012)\citenamefont
  {Parkinson}, \citenamefont {Manz}, \citenamefont {Novotn\'{y}}, \citenamefont
  {Sprunger}, \citenamefont {Kurtz}, \citenamefont {Schmid}, \citenamefont
  {Sholl},\ and\ \citenamefont {Diebold}}]{GarethPRB2012}%
  \BibitemOpen
  \bibfield  {author} {\bibinfo {author} {\bibfnamefont {G.~S.}\ \bibnamefont
  {Parkinson}}, \bibinfo {author} {\bibfnamefont {T.~A.}\ \bibnamefont {Manz}},
  \bibinfo {author} {\bibfnamefont {Z.}~\bibnamefont {Novotn\'{y}}}, \bibinfo
  {author} {\bibfnamefont {P.~T.}\ \bibnamefont {Sprunger}}, \bibinfo {author}
  {\bibfnamefont {R.~L.}\ \bibnamefont {Kurtz}}, \bibinfo {author}
  {\bibfnamefont {M.}~\bibnamefont {Schmid}}, \bibinfo {author} {\bibfnamefont
  {D.~S.}\ \bibnamefont {Sholl}}, \ and\ \bibinfo {author} {\bibfnamefont
  {U.}~\bibnamefont {Diebold}},\ }\href {\doibase 10.1103/PhysRevB.85.195450}
  {\bibfield  {journal} {\bibinfo  {journal} {Phys. Rev. B}\ }\textbf {\bibinfo
  {volume} {85}},\ \bibinfo {pages} {195450} (\bibinfo {year}
  {2012})}\BibitemShut {NoStop}%
\bibitem [{\citenamefont {Mulakaluri}\ \emph {et~al.}(2009)\citenamefont
  {Mulakaluri}, \citenamefont {Pentcheva}, \citenamefont {Wieland},
  \citenamefont {Moritz},\ and\ \citenamefont {Scheffler}}]{MulakaluriPRL2009}%
  \BibitemOpen
  \bibfield  {author} {\bibinfo {author} {\bibfnamefont {N.}~\bibnamefont
  {Mulakaluri}}, \bibinfo {author} {\bibfnamefont {R.}~\bibnamefont
  {Pentcheva}}, \bibinfo {author} {\bibfnamefont {M.}~\bibnamefont {Wieland}},
  \bibinfo {author} {\bibfnamefont {W.}~\bibnamefont {Moritz}}, \ and\ \bibinfo
  {author} {\bibfnamefont {M.}~\bibnamefont {Scheffler}},\ }\href {\doibase
  10.1103/PhysRevLett.103.176102} {\bibfield  {journal} {\bibinfo  {journal}
  {Phys. Rev. Lett.}\ }\textbf {\bibinfo {volume} {103}},\ \bibinfo {pages}
  {176102} (\bibinfo {year} {2009})}\BibitemShut {NoStop}%
\bibitem [{\citenamefont {McCarty}\ and\ \citenamefont
  {Bartelt}(2003)}]{McCartySS2003}%
  \BibitemOpen
  \bibfield  {author} {\bibinfo {author} {\bibfnamefont {K.~F.}\ \bibnamefont
  {McCarty}}\ and\ \bibinfo {author} {\bibfnamefont {N.~C.}\ \bibnamefont
  {Bartelt}},\ }\href {\doibase
  http://dx.doi.org/10.1016/S0039-6028(03)00003-7} {\bibfield  {journal}
  {\bibinfo  {journal} {Surf. Sci.}\ }\textbf {\bibinfo {volume} {527}},\
  \bibinfo {pages} {L203 } (\bibinfo {year} {2003})}\BibitemShut {NoStop}%
\bibitem [{Note1()}]{Note1}%
  \BibitemOpen
  \bibinfo {note} {The Verwey transition is first order for stoichimetric
  samples, while it is second order for extreme deviations of the composition
  (R. Arag\'{o}n, D.J. Buttrey, J.P. Shepherd, J.M. Honig, Phys. Rev. B
  {\protect \bf 31}, 430 (1985))}\BibitemShut {NoStop}%
\bibitem [{\citenamefont {Walz}(2002)}]{WalzJPc2002}%
  \BibitemOpen
  \bibfield  {author} {\bibinfo {author} {\bibfnamefont {F.}~\bibnamefont
  {Walz}},\ }\href {http://www.iop.org/EJ/abstract/0953-8984/14/12/203}
  {\bibfield  {journal} {\bibinfo  {journal} {J. Phys. Cond. Mat.}\ }\textbf
  {\bibinfo {volume} {14}},\ \bibinfo {pages} {R285} (\bibinfo {year}
  {2002})}\BibitemShut {NoStop}%
\bibitem [{\citenamefont {{Garc\'{\i}a}}\ and\ \citenamefont
  {{Sub\'{\i}as}}(2004)}]{GarciaJPc2004}%
  \BibitemOpen
  \bibfield  {author} {\bibinfo {author} {\bibfnamefont {J.}~\bibnamefont
  {{Garc\'{\i}a}}}\ and\ \bibinfo {author} {\bibfnamefont {G.}~\bibnamefont
  {{Sub\'{\i}as}}},\ }\href
  {http://www.iop.org/EJ/abstract/0953-8984/16/7/R01/} {\bibfield  {journal}
  {\bibinfo  {journal} {J. Phys. Cond. Mat.}\ }\textbf {\bibinfo {volume}
  {16}},\ \bibinfo {pages} {R145} (\bibinfo {year} {2004})}\BibitemShut
  {NoStop}%
\bibitem [{\citenamefont {Cornell}\ and\ \citenamefont
  {Schwertmann}(1997)}]{Cornellbook}%
  \BibitemOpen
  \bibfield  {author} {\bibinfo {author} {\bibfnamefont {R.~M.}\ \bibnamefont
  {Cornell}}\ and\ \bibinfo {author} {\bibfnamefont {U.}~\bibnamefont
  {Schwertmann}},\ }\href@noop {} {\emph {\bibinfo {title} {The Iron Oxides}}}\
  (\bibinfo  {publisher} {John Wiley \& Sons Ltd},\ \bibinfo {year} {1997})\
  p.\ \bibinfo {pages} {604}\BibitemShut {NoStop}%
\bibitem [{\citenamefont {Sanz}\ \emph {et~al.}(2013)\citenamefont {Sanz},
  \citenamefont {Oujja}, \citenamefont {Rebollar}, \citenamefont {Marco},
  \citenamefont {de~la Figuera}, \citenamefont {Monti}, \citenamefont
  {Bollero}, \citenamefont {Camarero}, \citenamefont {Pedrosa}, \citenamefont
  {Garc\'{\i}a-Hern\'{a}ndez},\ and\ \citenamefont
  {Castillejo}}]{MikelASS2013}%
  \BibitemOpen
  \bibfield  {author} {\bibinfo {author} {\bibfnamefont {M.}~\bibnamefont
  {Sanz}}, \bibinfo {author} {\bibfnamefont {M.}~\bibnamefont {Oujja}},
  \bibinfo {author} {\bibfnamefont {E.}~\bibnamefont {Rebollar}}, \bibinfo
  {author} {\bibfnamefont {J.~F.}\ \bibnamefont {Marco}}, \bibinfo {author}
  {\bibfnamefont {J.}~\bibnamefont {de~la Figuera}}, \bibinfo {author}
  {\bibfnamefont {M.}~\bibnamefont {Monti}}, \bibinfo {author} {\bibfnamefont
  {A.}~\bibnamefont {Bollero}}, \bibinfo {author} {\bibfnamefont
  {J.}~\bibnamefont {Camarero}}, \bibinfo {author} {\bibfnamefont {F.~J.}\
  \bibnamefont {Pedrosa}}, \bibinfo {author} {\bibfnamefont {M.}~\bibnamefont
  {Garc\'{\i}a-Hern\'{a}ndez}}, \ and\ \bibinfo {author} {\bibfnamefont
  {M.}~\bibnamefont {Castillejo}},\ }\href@noop {} {\bibfield  {journal}
  {\bibinfo  {journal} {Appl. Surf. Sci.}\ }\textbf {\bibinfo {volume} {282}},\
  \bibinfo {pages} {642} (\bibinfo {year} {2013})}\BibitemShut {NoStop}%
\bibitem [{\citenamefont {Altman}(2010)}]{altman_trends_2010}%
  \BibitemOpen
  \bibfield  {author} {\bibinfo {author} {\bibfnamefont {M.~S.}\ \bibnamefont
  {Altman}},\ }\href {http://stacks.iop.org/0953-8984/22/i=8/a=084017}
  {\bibfield  {journal} {\bibinfo  {journal} {J. Phys. Cond. Mat.}\ }\textbf
  {\bibinfo {volume} {22}},\ \bibinfo {pages} {084017} (\bibinfo {year}
  {2010})}\BibitemShut {NoStop}%
\bibitem [{\citenamefont {de~la Figuera}\ \emph {et~al.}(2006)\citenamefont
  {de~la Figuera}, \citenamefont {Puerta}, \citenamefont {Cerda}, \citenamefont
  {Gabaly},\ and\ \citenamefont
  {McCarty}}]{de_la_figuera_determiningstructure_2006}%
  \BibitemOpen
  \bibfield  {author} {\bibinfo {author} {\bibfnamefont {J.}~\bibnamefont
  {de~la Figuera}}, \bibinfo {author} {\bibfnamefont {J.~M.}\ \bibnamefont
  {Puerta}}, \bibinfo {author} {\bibfnamefont {J.~I.}\ \bibnamefont {Cerda}},
  \bibinfo {author} {\bibfnamefont {F.~E.}\ \bibnamefont {Gabaly}}, \ and\
  \bibinfo {author} {\bibfnamefont {K.~F.}\ \bibnamefont {McCarty}},\ }\href
  {\doibase http://dx.doi.org/10.1016/j.susc.2006.02.027} {\bibfield  {journal}
  {\bibinfo  {journal} {Surf. Sci.}\ }\textbf {\bibinfo {volume} {600}},\
  \bibinfo {pages} {L105 } (\bibinfo {year} {2006})}\BibitemShut {NoStop}%
\bibitem [{\citenamefont {McCarty}\ and\ \citenamefont {de~la
  Figuera}(2013)}]{KevinSSbook}%
  \BibitemOpen
  \bibfield  {author} {\bibinfo {author} {\bibfnamefont {K.~F.}\ \bibnamefont
  {McCarty}}\ and\ \bibinfo {author} {\bibfnamefont {J.}~\bibnamefont {de~la
  Figuera}},\ }in\ \href
  {http://www.springer.com/physics/condensed+matter+physics/book/978-3-642-34242-4}
  {\emph {\bibinfo {booktitle} {Surface Science Techniques}}},\ \bibinfo
  {series} {Springer Series in Surface Sciences}, Vol.~\bibinfo {volume} {51}\
  (\bibinfo  {publisher} {Springer Berlin Heidelberg},\ \bibinfo {year}
  {2013})\ p.\ \bibinfo {pages} {531}\BibitemShut {NoStop}%
\bibitem [{\citenamefont {Nie}\ \emph {et~al.}(2013)\citenamefont {Nie},
  \citenamefont {Starodub}, \citenamefont {Monti}, \citenamefont {Siegel},
  \citenamefont {Vergara}, \citenamefont {El~Gabaly}, \citenamefont {Bartelt},
  \citenamefont {de~la Figuera},\ and\ \citenamefont {McCarty}}]{niejacs2013}%
  \BibitemOpen
  \bibfield  {author} {\bibinfo {author} {\bibfnamefont {S.}~\bibnamefont
  {Nie}}, \bibinfo {author} {\bibfnamefont {E.}~\bibnamefont {Starodub}},
  \bibinfo {author} {\bibfnamefont {M.}~\bibnamefont {Monti}}, \bibinfo
  {author} {\bibfnamefont {D.~A.}\ \bibnamefont {Siegel}}, \bibinfo {author}
  {\bibfnamefont {L.}~\bibnamefont {Vergara}}, \bibinfo {author} {\bibfnamefont
  {F.}~\bibnamefont {El~Gabaly}}, \bibinfo {author} {\bibfnamefont {N.~C.}\
  \bibnamefont {Bartelt}}, \bibinfo {author} {\bibfnamefont {J.}~\bibnamefont
  {de~la Figuera}}, \ and\ \bibinfo {author} {\bibfnamefont {K.~F.}\
  \bibnamefont {McCarty}},\ }\href {\doibase 10.1021/ja402599t} {\bibfield
  {journal} {\bibinfo  {journal} {J. Am. Chem. Soc.}\ }\textbf {\bibinfo
  {volume} {135}},\ \bibinfo {pages} {10091} (\bibinfo {year}
  {2013})}\BibitemShut {NoStop}%
\bibitem [{\citenamefont {de~la Figuera}\ \emph {et~al.}(2013)\citenamefont
  {de~la Figuera}, \citenamefont {Vergara}, \citenamefont {N'Diaye},
  \citenamefont {Quesada},\ and\ \citenamefont
  {Schmid}}]{delaFigueraUltra2013}%
  \BibitemOpen
  \bibfield  {author} {\bibinfo {author} {\bibfnamefont {J.}~\bibnamefont
  {de~la Figuera}}, \bibinfo {author} {\bibfnamefont {L.}~\bibnamefont
  {Vergara}}, \bibinfo {author} {\bibfnamefont {A.~T.}\ \bibnamefont
  {N'Diaye}}, \bibinfo {author} {\bibfnamefont {A.}~\bibnamefont {Quesada}}, \
  and\ \bibinfo {author} {\bibfnamefont {A.~K.}\ \bibnamefont {Schmid}},\
  }\href {\doibase http://dx.doi.org/10.1016/j.ultramic.2013.02.020} {\bibfield
   {journal} {\bibinfo  {journal} {Ultramicroscopy}\ }\textbf {\bibinfo
  {volume} {130}},\ \bibinfo {pages} {77 } (\bibinfo {year}
  {2013})}\BibitemShut {NoStop}%
\bibitem [{\citenamefont {Roelofs}(1996)}]{RoelofsHSS1996}%
  \BibitemOpen
  \bibfield  {author} {\bibinfo {author} {\bibfnamefont {L.}~\bibnamefont
  {Roelofs}},\ }in\ \href
  {http://www.sciencedirect.com/science/article/pii/S1573433196800187} {\emph
  {\bibinfo {booktitle} {Handbook of Surface Science}}},\ Vol.~\bibinfo
  {volume} {1},\ \bibinfo {editor} {edited by\ \bibinfo {editor} {\bibfnamefont
  {W.}~\bibnamefont {Unertl}}}\ (\bibinfo  {publisher} {North-Holland},\
  \bibinfo {year} {1996})\ pp.\ \bibinfo {pages} {713--807}\BibitemShut
  {NoStop}%
\bibitem [{Note2()}]{Note2}%
  \BibitemOpen
  \bibinfo {note} {Lack of data very near to $T_c$ precludes making useful
  estimates of critical exponents.}\BibitemShut {Stop}%
\bibitem [{\citenamefont {Tromp}\ \emph {et~al.}(1996)\citenamefont {Tromp},
  \citenamefont {Theis},\ and\ \citenamefont {Bartelt}}]{trompprl1996}%
  \BibitemOpen
  \bibfield  {author} {\bibinfo {author} {\bibfnamefont {R.~M.}\ \bibnamefont
  {Tromp}}, \bibinfo {author} {\bibfnamefont {W.}~\bibnamefont {Theis}}, \ and\
  \bibinfo {author} {\bibfnamefont {N.~C.}\ \bibnamefont {Bartelt}},\ }\href
  {\doibase 10.1103/PhysRevLett.77.2522} {\bibfield  {journal} {\bibinfo
  {journal} {Phys. Rev. Lett.}\ }\textbf {\bibinfo {volume} {77}},\ \bibinfo
  {pages} {2522} (\bibinfo {year} {1996})}\BibitemShut {NoStop}%
\bibitem [{\citenamefont {Kinzel}\ and\ \citenamefont
  {Schick}(1981)}]{kinzel1981}%
  \BibitemOpen
  \bibfield  {author} {\bibinfo {author} {\bibfnamefont {W.}~\bibnamefont
  {Kinzel}}\ and\ \bibinfo {author} {\bibfnamefont {M.}~\bibnamefont
  {Schick}},\ }\href {\doibase 10.1103/PhysRevB.24.324} {\bibfield  {journal}
  {\bibinfo  {journal} {Phys. Rev. B}\ }\textbf {\bibinfo {volume} {24}},\
  \bibinfo {pages} {324} (\bibinfo {year} {1981})}\BibitemShut {NoStop}%
\bibitem [{Note3()}]{Note3}%
  \BibitemOpen
  \bibinfo {note} {The complications of enforcing charge states in this system
  make theoretical estimates difficult. Calculations in Ref. [\protect
  \rev@citealpnum {GarethPRB2012}] suggest that replacing one Fe$^{3+}$ with
  Fe$^{2+}$ costs 55 meV, which would be of the same order as we predict, but
  since the details of the configuration studied is not given, it is difficult
  to directly compare with our model.}\BibitemShut {Stop}%
\bibitem [{Note4()}]{Note4}%
  \BibitemOpen
  \bibinfo {note} {I. Bernal and S. Gallego, to be published.}\BibitemShut
  {Stop}%
\bibitem [{\citenamefont {Novotny}\ \emph {et~al.}(2013)\citenamefont
  {Novotny}, \citenamefont {Mulakaluri}, \citenamefont {Edes}, \citenamefont
  {Schmid}, \citenamefont {Pentcheva}, \citenamefont {Diebold},\ and\
  \citenamefont {Parkinson}}]{ParkinsonPRB2013}%
  \BibitemOpen
  \bibfield  {author} {\bibinfo {author} {\bibfnamefont {Z.}~\bibnamefont
  {Novotny}}, \bibinfo {author} {\bibfnamefont {N.}~\bibnamefont {Mulakaluri}},
  \bibinfo {author} {\bibfnamefont {Z.}~\bibnamefont {Edes}}, \bibinfo {author}
  {\bibfnamefont {M.}~\bibnamefont {Schmid}}, \bibinfo {author} {\bibfnamefont
  {R.}~\bibnamefont {Pentcheva}}, \bibinfo {author} {\bibfnamefont
  {U.}~\bibnamefont {Diebold}}, \ and\ \bibinfo {author} {\bibfnamefont
  {G.~S.}\ \bibnamefont {Parkinson}},\ }\href {\doibase
  10.1103/PhysRevB.87.195410} {\bibfield  {journal} {\bibinfo  {journal} {Phys.
  Rev. B}\ }\textbf {\bibinfo {volume} {87}},\ \bibinfo {pages} {195410}
  (\bibinfo {year} {2013})}\BibitemShut {NoStop}%
\bibitem [{\citenamefont {Senn}\ \emph {et~al.}(2012)\citenamefont {Senn},
  \citenamefont {Wright},\ and\ \citenamefont {Attfield}}]{SennNature2012}%
  \BibitemOpen
  \bibfield  {author} {\bibinfo {author} {\bibfnamefont {M.~S.}\ \bibnamefont
  {Senn}}, \bibinfo {author} {\bibfnamefont {J.~P.}\ \bibnamefont {Wright}}, \
  and\ \bibinfo {author} {\bibfnamefont {J.~P.}\ \bibnamefont {Attfield}},\
  }\href@noop {} {\bibfield  {journal} {\bibinfo  {journal} {Nature}\ }\textbf
  {\bibinfo {volume} {481}},\ \bibinfo {pages} {173} (\bibinfo {year}
  {2012})}\BibitemShut {NoStop}%
\end{thebibliography}%

\end{document}